\documentstyle[preprint,aps,epsf,prc]{revtex}
\tightenlines

\begin{document}
\title{The $\eta_c\gamma \gamma ^*$ transition form factor in the decay 
$\eta_c\to \gamma \ell^+\ell^-$ and in the crossed channels 
$\gamma e^-\to \eta_c e^- $ and $e^+e^-\to\eta_c\gamma.$}
\author{Elvio Di Salvo}
\address{\it Dipartimento di Fisica dell' Universit\`a di Genova and I. N. F. N.
- Sez. Genova, Via Dodecaneso 33, 16146 Genova, Italy}
\author{Michail P. Rekalo \footnote{ Permanent address:
\it National Science Center KFTI, 310108 Kharkov, Ukraine}
}
\address{Middle East Technical University, 
Physics Department, Ankara 06531, Turkey}
\author{Egle Tomasi-Gustafsson}
\address{\it DAPNIA/SPhN, CEA/Saclay, 91191 Gif-sur-Yvette Cedex, 
France}

\date{\today}
\maketitle
\begin{abstract}
The Dalitz distribution and the effective mass spectrum of the lepton pair
produced in the decay $\eta_c\to \gamma \ell^+ \ell^-$, as well as the 
differential cross sections for the crossed channels $\gamma e^-\to \eta_c e^- 
$ and $e^+ e^-\to \eta_c \gamma$, are calculated in terms of the sole 
electromagnetic transition form factor for the vertex $\eta_c\to\gamma 
\gamma^*$, where $\gamma^*$ is a virtual photon with a space-like (for $\gamma 
e^- \to \eta_c e^- $) or a time-like (for  $e^+ e^-\to \eta_c \gamma$ and 
$\eta_c\to\gamma \ell^+\ell^-$) four-momentum. Explicit parametrisations of the 
transition form factor are found in two different approaches, namely in Vector 
Dominance Model and in a QCD-inspired consideration on the basis of the process 
$c \overline{c}\to \gamma \ell^+ \ell^-$, with zero relative three-momentum of 
the annihilating quarks. 
\end{abstract}
\section{Introduction}

The radiative decays of mesons and baryons are an interesting source of 
important information for hadron spectroscopy and hadron 
electrodynamics\cite{fr,ap}. 
For example, it is worth recalling the role of such decays in determining the 
charmonium spectrum\cite{ap}. Among radiative decays, especially intriguing 
are those where lepton pairs are produced. A classical example is the 
process $\pi^0\to \gamma e^+ e^-$ \cite{Fo89}, which is sensitive to the radius 
of the 
corresponding form factor. Such a decay is typical of any neutral pseudoscalar 
meson $P$
(= $\eta,~\eta^\prime,~\eta_c$, etc.):
$$P\to\gamma \ell^+ \ell^-,~~\ell = e, \mu.$$
The corresponding transition electromagnetic form factors of
$P\to\gamma\gamma^*$-decay are complex functions of $t$, the four-momentum 
transfer squared, in the time-like region $4m_{\ell}^2 < t\le m^2$, where $m$ 
($m_{\ell}$) is the mass of $P$ ($\ell$). In particular  
the decay $\eta_c\to \gamma \ell^+ \ell^-$ offers the possibility of studying 
the transition form factor in a relatively large region of time-like momentum 
transfer,
$$4m_{\ell}^2\le t\le 9~ \mbox{GeV}^2.$$
A strong theoretical interest in the transition form factors of pseudoscalar 
mesons was stimulated by the experimental results of the CLEO collaboration 
\cite{Cl95,Cl97}, 
which measured these form factors at space-like momentum transfers, using the 
well known 2$\gamma$-exchange mechanism in the reaction 
$e^+ e^-\to e^+ e^- P$ (Fig. 1).
If one lepton, say the positron, is not detected, we can study the process of 
photoproduction of pseudoscalar mesons on electrons, $\gamma e^-\to 
P e^- $, in an unusual kinematics of colliding particles, $\gamma$ 
and $e^-$.
In the case of $\pi^0$ production the corresponding form  factor was measured 
up to $t=-8$ GeV$^2$, where, in principle, perturbative QCD 
applies\cite{Ja96,Kr96,Mu97,Bo89}.  Recently also the $\eta_c \gamma$ 
transition form factor has been determined\cite{L3} for two different values of 
$t$ in the spacelike region.

A special interest has the decay $\eta_c\to\gamma \ell^+ \ell^-$, as a 
generalisation of the decay $\eta_c\to \gamma \gamma$, whose  branching 
ratio is known\cite{pdg,L3} (see also predictions by dispersive 
methods\cite{svz,rry,ch,rry2,dp}). The
availability of a form factor instead of a single constant, as in 
the case of $\eta_c\to 2 \gamma$ decay, allows a better insight of the 
structure of the $\eta_c$-meson and of perturbative QCD. It is also interesting 
to look at the predictions of the "old" 
approach of Vector Dominance Model (VDM) in the unusual region of charmed 
particle decays\cite{Al94}.

This paper is devoted to a global approach to the study 
of all possible crossed channels, i. e.,
$$\eta_c\to \gamma \ell^+ \ell^-, ~~~~ \gamma e^-\to\eta_c e^-, ~~~~
e^+ e^-\to \eta_c \gamma,  $$
whose characteristics are driven by a single transition form factor. The 
numerical 
estimates are performed both within VDM and QCD, considering, in the latter 
case, the process $c \overline{c}\to \gamma \ell^+ \ell^-$ as the basic 
mechanism generating the $\eta_c\to \gamma \gamma*$ transition form factor.

Section II is dedicated to the Dalitz plot distribution of the decay
$\eta_c\to \gamma \ell^+ \ell^-$. In sect. III we propose two models for 
calculating the transition form factor $\eta_c \to \gamma \gamma*$, one based on 
VDM 
and the other on QCD. In sect. IV we study the reaction $e^+ e^-\to \eta_c 
\gamma$. Sect. V is dedicated to the process $\gamma e^-\to\eta_c e^-$. Lastly 
in sect. VI we draw some conclusions.

\section{The decay $\lowercase{{\eta}_c}\to\gamma\lowercase{\ell^+}
\lowercase{\ell^-}$.}
We start by considering the more general case of the decay of a meson
$M$ into a real $\gamma$ and a lepton pair. The differential decay width reads
\begin{equation}
d\Gamma=(2\pi)^4
\displaystyle\frac{\overline{|{\cal M}|^2}}{2m}
\delta^4(p-q-k_1-k_2)\displaystyle\frac{d^3q}{(2\pi)^32\omega}
\displaystyle\frac{d^3k_1}{(2\pi)^3 2E_1}
\displaystyle\frac{d^3k_2}{(2\pi)^3 2E_2},
\label{dalitz11}
\end{equation}
where $\omega$, $E_1$ and $E_2$ are the energies of 
the final $\gamma$, $e^+$ and $e^-$ respectively.

The notation of the  four-momenta is shown in Fig. 2; the line above the 
modulus squared of the matrix element $|{\cal M}|^2$ denotes the sum over the 
polarisations of all the particles produced in the decay.

The standard form of the matrix element ${\cal M}$ for the 
$M\to\gamma \ell^+ \ell^-$ decay in one-photon approximation is
$$ {\cal M}=\displaystyle\frac{e^2}{t}\ell_{\mu}\cal J^{\mu},$$
where $t=(k_1+k_2)^2$ is the effective mass squared of the leptonic pair,
$\ell_\mu=\overline{u}(k_2)\gamma_{\mu}u(-k_1)$ and $\cal J_{\mu}$
is the electromagnetic current in the decay  $M\to\gamma\gamma^*$ 
\footnote{In the decay considered here, $\cal J_{\mu}$ is 
proportional to the electric charge $e$.}. 

According to the usual notation we obtain
\begin{equation}
\overline{|{\cal M}|^2} = \frac{e^4}{t^2}L_{\mu\nu} W^{\mu\nu},
\label{modsq}
\end{equation}
where
\begin{equation}
L_{\mu\nu}=\overline{\ell_{\mu}\ell^*_{\nu}}, \ ~~~~~ \ W^{\mu\nu}=
\overline{\cal J^\mu\cal J^{\nu *}}.
\label{tensor}
\end{equation}
Substituting (\ref{modsq}) into (\ref{dalitz11}), and integrating over all 
variables except $E_{1,2}$, we get the two-lepton Dalitz distribution:
$$
d\Gamma=\displaystyle\frac{\alpha^2}{4\pi}
\displaystyle\frac{L_{\mu\nu}W^{\mu\nu}}{t^2m}dE_1dE_2,
$$ 
where $\alpha$ = $e^2/4\pi$.
However for our aims it is more convenient to express $E_1$ and $E_2$ in terms 
of the dimensionless variables $x=t/m^2$ and  $y=2(E_1-E_2)/m$ \cite{Du96},
which are defined in the ranges $\delta\le x\le 1$  and $-y_0\le y\le y_0$, 
$y_0=(1-x)\sqrt{1-{\delta}/{x}}$ and $\delta=4m_{\ell}^2/m^2$.
Furthermore we take into account, wherever possible, symmetry properties:
\begin{itemize}
\item Gauge invariance - $k\cdot\ell$=$k\cdot\cal J$=$0$ - implies 
that the product $L_{\mu\nu}W^{\mu\nu}$ can be 
rewritten in terms of the space components of the two tensors above, i. e.,
\begin{equation}
L_{\mu\nu}W^{\mu\nu}=(L_{xx}+L_{yy})W_{xx}+\displaystyle\frac{t^2}{k_0^4}
L_{zz}W_{zz},
\label{eq1}
\end{equation}
where $k_o$ denotes the energy of the $\gamma^*$. 
Here we have adopted a reference frame in which the meson $M$ is at rest and 
the $z-$axis is parallel to the momentum $\vec k $ of the virtual photon.
\item Parity invariance allows to write
$$W_{ij}=(\delta_{ij}-\hat{k_i}\hat{k_j})W_1(t)+\hat{k_i}\hat{k_j}W_2(t), \ 
~~~~ \ \hat{\vec k }=\vec k /|\vec k |,$$
where $W_{1(2)}(t)$ is the real structure function (SF) which 
describes the production of $\gamma^*$ with transverse (longitudinal) 
polarisation.
\end{itemize}
Taking into account all that, calculations yield
$$d\Gamma=\displaystyle\frac{\alpha^2}{16\pi}W_{1}(t)
\displaystyle\frac{dxdy}{xm}\left [1+\displaystyle\frac{4m_{\ell}^2}{t}+R_L(t)
\displaystyle\frac{t}{k_0^2}+y^2\displaystyle\frac{m^2}{4\vec k ^2}\left 
(1-R_L(t)
\displaystyle\frac{t}{k_0^2}\right )\right ],$$
\begin{equation}
R_L(t)=W_2(t)/W_1(t),~~k_0=\displaystyle\frac{m^2+t}{2m},~~
\vec k^2=\displaystyle\frac{(m^2-t)^2}{4m^2}.
\label{eq2}
\end{equation}

One can see that the differential width, which characterises the Dalitz 
distribution for the decay $M\to \gamma \ell^+\ell^-$, is symmetric 
with respect to the exchange $E_1\leftrightarrow E_2$, in agreement with 
$C$-invariance of the electromagnetic interaction.

The specific dependence of $d^2\Gamma/dxdy$ on the variable $y$,
$$d^2\Gamma/dxdy=a(x)+y^2b(x),$$
is a direct consequence of one-photon mechanism. It has the 
same origin as the $\cot^2\frac{\theta_e}{2}$ dependence of the 
differential cross section in electron-hadron scattering (where $\theta_e$ is 
the electron scattering angle in the laboratory system) and as the 
$\cos^2\theta$-dependence of the differential cross section for the inclusive 
process 
$e^+e^-\to h\ X$, where $h$ denotes the final detected hadron and $\theta$ is 
the angle between the three-momenta of the hadron and of the electron  in the 
overall CMS.

One can see that the value $W_1(0)$ determines the probability of the radiative 
decay width $\Gamma_0 \equiv \Gamma(M\to 2\gamma)$, i.e., 
$$\Gamma_0=\displaystyle\frac{1}{2}
(2\pi)^4e^2\displaystyle\frac{W_{xx}+W_{yy}}{2m}
\int\delta^4(p-q-k)\displaystyle\frac{d^3k}{(2\pi)^32\omega_1}
\displaystyle\frac{d^3q}{(2\pi)^3 
2\omega_2}=\displaystyle\frac{\alpha}{4m}W_1(0).$$
This allows to rewrite the Dalitz distribution in the following form:
\begin{equation}
\displaystyle\frac{1}{\Gamma_0}
\displaystyle\frac{d^2\Gamma}{dxdy}(M\to \gamma\ell^+\ell^-)
=\displaystyle\frac{\alpha}{4\pi}R_T(t)\displaystyle\frac{1}{x}
\left [ 1+\displaystyle\frac{4m^2_{\ell}}{t}+R_L(t)
\displaystyle\frac{t}{k_0^2}+
\displaystyle\frac{y^2}{4\vec k ^2}m^2
\left (1-R_L(t)\displaystyle\frac{t}{k_0^2}\right )\right],
\label{eq3}
\end{equation}
where $R_T(t)=W_1(t)/W_1(0)$ and $R_L(t)$ is defined in  Eq. (\ref{eq2}).

Therefore the study of the energy distribution of leptons in $M\to \gamma 
\ell^+\ell^-$ in the Dalitz plane allows to determine $R_L(t)$ and $R_T(t)$, 
i. e., the two fundamental quantities of the decay considered,  as 
functions of the effective mass of the $\ell^+\ell^-$-system.

The result of the integration in Eq. (\ref{eq3}) over the variable $y$ 
determines the $\ell^+\ell^-$ effective mass spectrum in the decay
$M\to \gamma\ell^+\ell^-$ \cite{Kr55}:
\begin{equation}
\displaystyle\frac{1}{\Gamma_0}
\displaystyle\frac{d\Gamma}
{dx}(M\to \gamma\ell^+\ell^-)=\displaystyle\frac{2\alpha}{3\pi}R_T(t)
\displaystyle\frac{(1-x)}{x}
\left (1+2\displaystyle\frac{m_{\ell}^2}{t}\right 
)\sqrt{1-\displaystyle\frac{4m^2_{\ell}}{t}}\left 
[1+R_L(t)\displaystyle\frac{t}{2k_0^2}\right ].
\end{equation}

Turning to the specific decay  $\eta_c \to \gamma\gamma^*$, in this case (and 
more generally for any neutral pseudoscalar meson) the hadron
electromagnetic current reads

\begin{equation}
{\cal J}_\mu=\displaystyle\frac{F(t)}{m}\epsilon_{\mu\alpha\beta\gamma}k^\alpha 
e^{\beta}q^{\gamma},
\label{etaci} 
\end{equation}
where $e_\beta$ is the polarisation four-vector of the real photon and $F(t)$ 
the electromagnetic form factor of the transition $\eta_c\to \gamma \gamma^*$, 
which has a nonzero imaginary part for $t\geq 4m_\pi^2$. Note that we have 
defined the form factor differently from Lepage and Brodsky\cite{Le80}.
Inserting Eq. (\ref{etaci}) into the definition of the tensor $W_{\mu\nu}$, 
Eq. (\ref{tensor}), we obtain
$$W_1(t)=|F(t)|^2(1-x)^2\displaystyle\frac{m^2}{4}, \ ~~~~ \ W_2(t)=0,$$
i.e. $R_L(t)=0.$

There are three different processes for studying the $t-$dependence of $F$:

\begin{enumerate}
\item
The photoproduction of $\eta_c$ in 
$\gamma e^- \to \eta_c e^- $ at space-like momentum transfers;
\item The decay $\eta_c\to \gamma \ell^+\ell^-$, at 
time-like momentum transfers, $ 4m_{\ell}^2\le t\le m^2$;
\item The reaction  $e^+e^-\to \eta_c\gamma $, at time-like momentum 
transfers, $ t\ge m^2$. 
\end{enumerate}
Therefore the simultaneous study of all these different processes can allow to 
determine the $t-$dependence of the electromagnetic form factor $F(t)$ 
for any value of $t$ (see Fig. 3).
Moreover all the possible observables in the above mentioned processes 
depend only on $|F(t)|^2$. This means that the phase of $F(t)$ in the time-like 
region cannot be measured. But the determination of this  phase as a 
function of $t$ is very important. For example, a dispersion relation for 
the form factor $F(t)$, considered as an analytical function of $t$, can be 
analysed only if both the modulus and the phase of $F(t)$ are known. The 
problem of measuring the phase of  complex form factors of hadrons 
in the time-like region is not yet solved. This is a common problem of  
the electromagnetic form factor of the charged pion, of the neutral and charged 
kaons, etc..

The determination of the transition form factor $F(t)$ is necessary for  
calculating the following observables:
\begin{itemize}
\item The Dalitz-distribution in the variables $x$ and $y$;
\item The spectrum of the effective masses in $\eta_c\to \gamma \ell^+\ell^-$;
\item The coefficient of internal conversion, i.e. the ratio of the decay width 
of $\eta_c\to \gamma\ell^+\ell^-$ to the decay width of $\eta_c\to 
\gamma\gamma$:
\begin{equation}
I=\displaystyle\frac{\Gamma(\eta_c\to \gamma \ell^+\ell^-)}
{\Gamma_0}=
\displaystyle\frac{2\alpha}{3\pi}
\int^1_{\delta} dx \left |
\displaystyle\frac{F(t)}{F(0)}
\right |^2\displaystyle\frac{(1-x)^3}{x}
\left (1+2\displaystyle\frac{m_{\ell}^2}{t}\right )
\left (1-\displaystyle\frac{4m_{\ell}^2}{t}\right )^{1/2}.
\end{equation}
\end{itemize}
This coefficient derives its main contribution from small values of $t$,
therefore it is sensitive to the mass of the lepton, in particular it is
different in
the decays $\eta_c\to \gamma e^+ e^-$ and $\eta_c\to \gamma \mu^+ \mu^-$.
\section{Two models for $F(\lowercase{t})$}
Now we consider the predictions of two models, the VDM and a pQCD-inspired 
non-relativistic model.

\subsection{Vector Dominance Model}

In the framework of VDM (Fig. 4), the $t-$dependence of the electromagnetic 
form factor $F(t)$ can be written as
\begin{equation} 
F(t)=\sum_V\displaystyle\frac{g_V}
{t-m^2_V}, \ ~~~~~ \ g_V=m^2_{V}f_Vg(V\to \eta_c\gamma),
\end{equation}
having summed over the contributions of the vector mesons. $m_V$ is the
vector meson mass, $f_V$ is the constant of the $\gamma^*\to V$-transition and 
$g(V\to \eta_c\gamma)$ is the constant in the vertex 
$V\to \eta_c\gamma$. If $m_V\ge m$, the decay $V\to 
\eta_c\gamma$ is possible and the corresponding branching ratio allows to 
determine the constant $g(V\to\eta_c\gamma)$, or, more exactly, as previously 
discussed, the modulus of this constant. Incidentally, notice that the constant 
$f_V$ determines the decay width of $V\to\ell^+ \ell^-$ (see Appendix).

The constants $g_V$ satisfy two important conditions:
\begin{eqnarray}
&-\sum_V\displaystyle\frac{g_V}{m^2_{V}}=F(0),
\label{r1}\\
&\sum_V{g_V}=0.
\label{r2}
\end{eqnarray}
The first one is a relation between different $g_V$, through an evident 
normalisation 
of $F(t)$. The constant $ F(0)$ can be determined  from the decay width of
$\eta_c\to 2\gamma$, i.e., from
\begin{equation}
\Gamma_0=\displaystyle\frac{\alpha}{4}
\left |F(0)\right |^2m,
\end{equation} 
which is experimentally known. Again, the value of the width $\Gamma_0$ does not 
fix the sign of $F(0)$, which introduces a two-fold ambiguity in relation 
(\ref{r1}).

Relation (\ref{r2}) results from a specific asymptotic behaviour of $F(t)$: the 
structure of the current ${\cal J}_\mu$ shows that 
$F(t) \propto t^{-2} $ for large $|t|$.  This differs from the standard 
$|t|^{-1}$ behaviour of the elastic electromagnetic form factors of mesons, 
however it agrees 
with the quark counting rule, and it results from the presence in Eq. (8) of 
an additional four-momentum, owing to the electromagnetic current of the 
decay considered.

The simplest model that can be suggested here for $F(t)$ must contain the 
contribution of two vector mesons. The problem is how to select the 
most important contributions.

Taking into account the $c\overline{c}$-nature of $\eta_c$, the $J/\psi$ seems 
to be a good candidate, as confirmed by the value of the branching ratio 
$Br(J/\psi\to \eta_c\gamma)=(1.3\pm 0.4)\%$. Similarly the branching ratio 
$Br(\eta_c\to \rho \rho)=(2.6\pm 0.9)\%$ exhibits
the importance of the $\rho$-contribution to the form factor $F(t)$. Therefore 
in such a two-pole model we find the following two parametrisations: 
\begin{equation}
F(t)=\pm |F(0)|\displaystyle\frac{m^2_{\rho}m^2_{J/\psi}}
{m^2_{J/\psi}-m^2_{\rho}}\left 
(\displaystyle\frac{1}{t-m^2_{\rho}}-\displaystyle\frac{1}{t-m^2_{J/\psi}}\right 
),
\label{pole}
\end{equation}
which cannot be disentangled in the study of the above mentioned processes.

In order to avoid singularities at $t$=$m_V^2$, we have to substitute
$m^2_V\to m^2_V-im_V\Gamma_V$, where $\Gamma_V$ is the width of the $V-$meson
($V$ = $\rho$, $J/\psi$).

These parametrisations offer a numerical estimation of all the observables of
the $\eta_c$ photoproduction on electrons, $\gamma e^-\to \eta_c e^- $, and of 
the decay 
$\eta_c\to \gamma \ell^+ \ell^-$, where $t\le m^2$. In both processes the 
$\rho$-contribution is particularly important at relatively small momentum 
transfers, $|t|\le 1$ GeV$^2$. The $J/\psi$-contribution plays an 
important role in the decay $\eta_c\to \gamma \ell^+ \ell^-$, near the upper 
kinematical limit in the variable $t$, $t\to m^2$, where the $J/\psi$-pole is 
very close to the physical region.

On the other hand this simplified parametrisation is not convenient for the  
study of the annihilation process, $e^+ e^-\to\eta_c \gamma$, to which the 
entire set of $\psi$-resonances in principle could contribute. In this case 
formula (\ref{pole}) has to be generalised, taking into account, at least,
\begin{itemize}
\item the contribution of additional $\psi$ resonances;
\item the widths of such resonances, as illustrated above.
\end{itemize}
In the framework of VDM it is in principle possible to write a five-pole 
representation for $F(t)$, with the contributions of 
$\rho,~J/\psi,~\psi(3.77),~\psi(4.04),~\psi(4.41).$
The constants $g_V$ of $\rho,~J/\psi$ and $~\psi(3.77)$ can be found using 
the existing experimental data about the radiative decays: $J/\psi\to\eta_c
\gamma$, $\psi(3.77)\to\eta_c \gamma$, $\eta_c\to\rho \rho$, $\rho\to 
e^+ e^-$, $J/\psi\to e^+ e^-$ and $\psi(3.77)\to e^+ e^-$. The details of 
the calculation are given in the Appendix. On the other hand, using relations 
(\ref{r1}) and 
(\ref{r2}), it is possible to find  the constants $g_V$ of the 
$\psi(4.04)$ and the $\psi(4.41)$-resonances. As
two choices are possible for the sign of $F(0)$ and for the constants $g_V$ 
for the first three resonances, $\rho,~J/\psi$ and $~\psi(3.77)$, one finds 
a  $2^4=16$-fold 
parametrisation for $F(t)$, which prevents any realistic 
estimation of the $t-$behaviour of the cross section for 
the process $e^+ e^-\to \eta_c \gamma$.
So our prediction of the various characteristics of the decay $\eta_c \to 
\gamma e^+ e^-$ will be based on the simplified two-pole model for $F(t)$.

\subsection{QCD-Inspired Model}

Now we consider the process of $c\overline{c}$-quark annihilation into 
$\gamma\gamma^*$,
which receives the contribution of the two diagrams illustrated in Fig. 5. 

In the limit of $|\vec p| = 0$ (where $\vec p$ is the 
relative momentum of the $c\overline{c}$-system in the CMS),  the 
$c\overline{c}$- annihilation in a singlet $S$-state - which has the quantum 
numbers of the $\eta_c$-meson, $J^P=0^-$ - is described by a matrix 
element of the form
\begin{equation}
{\cal M}(c\overline{c})\propto \vec k \cdot \vec e\times \vec {e'},
\label{pedro}
\end{equation}
where $\vec {e}$ ($\vec {e'}$) is the polarisation vector of the real (virtual) 
photon. The following relations hold: $\vec{e'}\cdot \vec k \ne 0$ and 
$\vec{e}\cdot \vec k = 0$. The result (\ref{pedro}) can be found, without 
specific calculations, on the basis of symmetry properties relative to $C-$ and 
$P-$transformations for the process $c\overline{c}\to \gamma\gamma^*$.
Due to the positive $C-$parity of the two final $\gamma$'s, the  sum of the 
total spin $S$ and the angular momentum $\ell$ of the ($c\overline{c}$)-system 
must be even. But in the limit of $\vec p\to 0$ we have $\ell$ = 0, therefore 
$S=0$. In this limit the $t-$dependence of the resulting 
matrix element for $c\overline{c}\to\gamma\gamma^*$ can be identified with 
the $t-$dependence of the form factor $F(t)$ of the decay 
$\eta_c\to \gamma\gamma^*$. This dependence can be written in terms of the 
$c-$quark propagators in the elementary process $c\overline{c}\to
\gamma\gamma^*$, which are identical for both diagrams of Fig. 5. Therefore in 
the limit of $\vec p=0$ we have
$$ F(t)\propto\displaystyle\frac{1}{(p-k)^2-m^2_c}=
\ \displaystyle\frac{1}{k^2-2p\cdot k} 
\stackrel{\vec p=0}{=} \displaystyle\frac{1}{k^2-2m_ck_0} 
=\displaystyle\frac{2}{k^2-4m_c^2},$$
which implies that $F(t)$ $\propto$ $|t|^{-1}$ for large $|t|$, unlike the VDM.

The parameter $4m^2_c$ which appears in the denominator
should be identified with a physical quantity. The simplest possibility is to 
set $2m_c=m_{J/\psi}$, which corresponds to the $J/\psi$-contribution to 
$F(t)$,  appearing in the framework of VDM.  A perturbative approach, on the 
basis of a factorisation of short- and long-distance physics \cite{Le80}, as
considered by Feldmann and Kroll\cite{Fe97,fe} in the space-like region, 
results in 
\begin{equation}
4m_c^2\to m^2+2<\vec k _{\perp}^2 >,
\label{recip}
\end{equation}
where $\vec k _\perp$ denotes the transverse momentum of the $c$-quark in the 
$\eta_c$-meson \cite{Wi85}. This substitution may be regarded as a possible 
procedure for taking into account the transverse momentum of the quark inside 
the meson. On the other hand, note that our QCD-inspired consideration above is 
valid for any 
$t$, both space-like and time-like, except for a small neighbourhood 
of $t$ = $4m_c^2$. Therefore, if we assume that the substitution (\ref{recip})
can be extended at least  up to $t \leq m^2$, $F(t)$ turns out to be very 
sensitive to $<\vec k ^2_\perp>$ near the upper kinematical limit of the Dalitz
plot, as results from Fig. 6. 
 
In this approach it is also possible to find the corresponding form factors of
the radiative decays $\chi_J \to \gamma \ell^+\ell^-$
 of the P-wave charmonium states, $J=0$, $1$ and $2$. To this end it is 
necessary to find the matrix element of the process $c  \overline{c} 
\to\gamma  \gamma^*$ (see Fig. 5). The spin structure of the $\chi$-decay
matrix elements is 
$$\xi_1^{\dagger}\sigma_2\vec\sigma\cdot\vec p\xi_2~\vec e\cdot\vec {e'}:
\ ~~~~~ \chi_0\to\gamma+\gamma^*,$$
$$\xi_1^{\dagger}\sigma_2~\vec\sigma\times\vec p\cdot\vec e\times
\vec k \xi_2~\vec {e'}\cdot\vec k : \ ~~~~~ \chi_1\to\gamma+\gamma^*,$$
$$\xi_1^{\dagger}\sigma_2(\sigma_a 
p_b+\sigma_bp_a-\displaystyle\frac{2}{3}\delta_{ab}\vec\sigma\cdot\vec p)\xi_2
\left \{ 
\begin{array}{cc}
e_{1a}&e_{1b}\\
k_ak_b&\vec {e}\cdot \vec {e'}\\
e_{1a}k_b&\vec {e'}\cdot \vec k 
\end{array}
\right \}: \ ~~~~~ \chi_2\to\gamma+\gamma^*,
$$
where  $\xi_{1(2)}$ is the two-component spinor of the $\overline{c}$ ($c$) 
quark.

In the general case the decays of $\chi_0$ and $\chi_1$ are characterised by 
one 
form factor, whereas in the case of $\chi_2\to\gamma\gamma^*$  three 
independent form 
factors are present. Notice also that the matrix element for 
$\chi_1\to\gamma\gamma^*$ is proportional to $\vec {e'}\cdot\vec k $, which 
vanishes for real photons, in agreement with the  Landau-Yang theorem.

As one can see from Fig. 6, the VDM and QCD-inspired model predict essentially 
different $t-$dependences for the transition form factor $F(t)$ in the whole 
region $4m_{\ell}^2\le t\le m^2$, where the decay $\eta_c\to \gamma \ell^+ 
\ell^-$ occurs. The coefficients of internal conversion of the decays 
$\eta_c\to \gamma e^+ e^-$ and 
$\eta_c\to \gamma \mu ^+ \mu^-$ 
are also strongly model-dependent (see Table 1).  These properties of $F(t)$ 
depend essentially on two factors: the role of the $\rho-$meson in the VDM 
approach and the small mass difference between $\eta_c$ and $J/\psi$.

Also in the space-like region the two models describe the form factor quite 
differently; in particular the QCD-inspired model predicts for $F(t)$ a much 
slower decrease at increasing $|t|$, in agreement with present data\cite{L3}.

\section{The reaction $\lowercase{e}^+ \lowercase{e}^-\to \lowercase{{\eta}_c}
~ \gamma$}
The differential cross section in the overall CMS reads
\begin{equation}
\displaystyle\frac{d\sigma}{d\Omega}=
\displaystyle\frac{ \overline{|{\cal M}|^2}}{64\pi^2t}
\displaystyle\frac{|\vec q|}{|\vec k |} =
\displaystyle\frac{\overline{|{\cal M}|^2}}{64\pi^2t}\left 
(1-\displaystyle\frac{m^2}{t}\right ), \ ~~~~ \ m_e=0,
\end{equation}
where $\vec k $ and $\vec q$ are the three-momenta of the colliding leptons and 
of the produced $\gamma$ and $t=(k_1+k_2)^2$ is the square of the total energy 
of the colliding particles. 

Taking into account Eq. (\ref{etaci}), we get
\begin{equation}
\displaystyle\frac{d\sigma}{d\Omega}=\displaystyle\frac{\alpha^2}{32m^2}(1+
\cos^2\theta)\left (1-\displaystyle\frac{m^2}{t}\right)^3|F(t)|^2, ~~ \ ~~~ \
\ ~~~ t\ge m^2.
\label{diff}
\end{equation}
The cross section for 
collinear kinematics (i.e. $\theta=0$ or $\pi$) is non-vanishing, in agreement
with helicity conservation. Indeed, in the limit of $m_e=0$, owing to the 
properties of the electromagnetic current $\overline{u}\gamma_{\mu}u$, we have 
$\lambda_i=\lambda_{e^-}+\lambda_{e^+}=\pm 1$; on the other hand for a real 
$\gamma$ we have $\lambda_f=\lambda_{\gamma}=\pm 1$.

In any process of the type $e^+ e^-\to \gamma^*\to h_1 \overline{h_2}$, where 
$h_1$ and $h_2$ are hadrons or hadronic states, the one-photon mechanism is 
characterised by a definite angular dependence:
$$\displaystyle\frac{d\sigma}{d\Omega}=A(t)+B(t)\cos^2\theta,$$
where $\theta$ is the production angle of $h_1$ in the overall CMS and 
$A(t)$ and $B(t)$ are two structure functions which are quadratic combinations 
of the electromagnetic 
form factors of the transition $\gamma^*\to h_1 \overline{h_2}$. In the 
case considered we have only one form factor for the vertex 
$\gamma^*\to \eta_c \gamma$, therefore $B(t)/A(t)=1$, as results from
the angular dependence (\ref{diff}).

The threshold behaviour of the cross section is proportional to $\left (1 
-\displaystyle\frac{m^2}{t}\right )^3$ and results from the magnetic dipole 
radiation in the transition $\gamma^*\to\eta_c \gamma$.

The factorisation of the $t-$ and $\cos\theta$-dependences in the differential 
cross section of $e^+ e^-\to \eta_c \gamma$, Eq. (\ref{diff}), is due to the 
specific structure of the matrix element and to the choice of the CMS as a 
reference frame. Therefore the $t-$dependence is the same for any production 
angle and coincides with the $t-$dependence of the total cross section:
\begin{equation}
\sigma=\displaystyle\frac{\pi}
{6m^2}\alpha^2\left(1-\displaystyle\frac{m^2}{t}\right )^3|F(t)|^2.
\label{tot}
\end{equation}
Formulae (\ref{diff}) and (\ref{tot}) can be re-written as
$$\displaystyle\frac{d\sigma}{d\Omega} = 
\displaystyle\frac{\alpha}{2 m^3}\Gamma_0 \left(1-
\displaystyle\frac{m^2}{t}\right )^3\left 
|\displaystyle\frac{F(t)}{F(0)}\right |^2 (1+\cos^2\theta) = 
\displaystyle\frac{
(1+\cos^2\theta)}{16\pi} 3 \sigma,$$
$$\sigma = \displaystyle\frac{4}{3}\displaystyle\frac
{\alpha\pi}{m^3}\Gamma_0\left(1-
\displaystyle\frac{m^2}{t}\right )^3\left 
|\displaystyle\frac{F(t)}{F(0)}\right |^2. $$
In order to predict the $t-$dependence of the total and differential 
cross sections, it is necessary to know the form factor $F(t)$ in the region 
$t\ge m^2$, and especially where the contribution of all the $\psi$-resonances 
is important. From our previous discussion it turns out that the VDM can be 
applied here. 
\section{The process  $\gamma \lowercase{e}^-\to \lowercase{{\eta}_c}
\lowercase{e}^- $.}
The direct study of the reaction $\gamma e^-\to \eta_c e^-$ is experimentally 
difficult due to the large threshold,
$E_{\gamma, thr}\simeq m^2/2m_e\simeq ~~ $9 TeV.
Therefore this process can be studied through the $2\gamma$ mechanism in the 
reaction $e^{\pm}e^-\to e^\pm e^-\eta_c$ (see Figs. 7 and 1) with a quasi-real 
photon produced in the $e^\pm\to e^\pm \gamma$-vertex. In this case, for the 
process $\gamma e^-\to \eta_ce^-$, it is necessary 
to consider a particular reference frame with colliding $\gamma$ and $e^-$. Let 
us re-write the cross section of $\gamma e^-\to \eta_ce^-$ in terms of the 
Mandelstam variables $s$, $t$ and $u$, which can be defined in the standard 
form, according to the following notation of the particles four-momenta:
$$\gamma(q)+e^-(k_1)\to e^-(k_2)+\eta_c(p),$$
$$s=(k_1+q)^2,$$
$$u=(k_2-q)^2,$$
$$t=(k_1-k_2)^2,$$
where $q$, $p$ and $k_1$ ($k_2$) are the four-momenta of the photon, of 
$\eta_c$ and of the initial (final) electron.
In the overall CMS the differential cross section can be written as
$$\displaystyle\frac{d\sigma}{d\Omega_{\eta}}(\gamma e^-\to \eta_ce^-) = 
\displaystyle\frac{\overline{|{\cal M}|^2}}{64\pi^2s} \left ( 
1-\displaystyle\frac{m^2}{s}\right ).
$$
In the one-photon exchange approximation
the expression of $|{\overline{\cal M}|^2}$ can be derived from  eq (5), by 
exploiting the crossing symmetry. In the direct channel we have
\begin{equation}
\overline{|{\cal M}|^2}=\displaystyle\frac{e^4  |F(t)|^2}{m^2 t}
(1+\cos^2\theta) (t-m^2)^2,
\end{equation}
where $t$ is the overall CMS energy squared.

Taking into account the relation 
$$1-\cos^2\theta=\displaystyle\frac{4su}{(t-m^2)^2},$$ 
the invariant differential cross section for $\gamma e^-\to e^-\eta_c$ results 
to be
\begin{equation}
\displaystyle\frac{d\sigma}{dt}=\displaystyle\frac{\pi\alpha^2}
{4s^2m^2|t|}\left [(t-m^2)^2-2us\right ] |F(t)|^2=
4\pi\displaystyle\frac{\alpha}{m^3}\Gamma_0\left | 
\displaystyle\frac{F(t)}{F(0)}\right 
|^2\displaystyle\frac{(t-m^2)^2-2us}{s^2|t|}.
\end{equation}

Since the four-momentum transfer in this process is space-like, the reaction 
considered is especially sensitive 
to the $\rho-$meson contribution, which is the lightest vector meson - in 
particular for $|t|\le 1$ GeV$^2$.

In the framework of one-photon mechanism a definite prediction for the 
polarisation observables can be derived. Independently of the kinematical 
conditions, the absorption asymmetry of a linearly polarised photon, i. e.,
$$\Sigma=
\displaystyle\frac{\sigma_{\perp}-\sigma_{\parallel}}
{\sigma_{\perp}+\sigma_{\parallel}},$$
where $\sigma_{\perp}$ ($\sigma_{\parallel}$) is the differential cross 
section with linearly polarised 
photons, whose polarisations are orthogonal (parallel) to the reaction plane - 
is equal to 1. This result, which implies $\sigma_{\parallel}$ = 0, follows 
from parity conservation.

This prediction can be tested using a linearly polarised photon beam and 
detecting the final electron. The electron polarisation here does not carry 
any additional information, owing to the presence of a single electromagnetic 
form factor. However for 
two-spin polarisation observables non-trivial effects can be produced by the 
collisions 
of circularly polarised photons with longitudinally polarised electrons.

\section{Conclusions}

We have considered some characteristics of the radiative decay 
$\eta_c\to\gamma \ell^+ \ell^-$, such as the Dalitz distribution with respect 
to the energies of the two final leptons, the effective mass spectrum of the 
lepton pair and the coefficient of internal 
conversion, i.e. the ratio  $\Gamma (\eta_c\to \gamma \ell^+ \ell^-)/
\Gamma(\eta_c\to 2\gamma)$. We have used two different models for the 
electromagnetic form factor $F(t)$. The VDM allows to find a relatively 
simple 2-pole representation for the form factor $F(t)$, with a definite 
normalisation at $t=0$ and asymptotic 
behaviour in $t^{-2}$, in agreement with the quark counting rule. 
The VDM parametrisation contains the contribution of the $\rho-$meson (which is 
important owing to the large width of the decay $\eta_c\to \rho^0\rho^0$) and 
the $J/\psi$-contribution.
This two-pole representation can be used for analysing the 
decay $\eta_c\to \gamma \ell^+ \ell^-$, which corresponds to timelike $t$, and 
the photoproduction process 
$\gamma e^-\to \eta_c e^- $, where $t$ is space-like. However 
in the calculation of the energy 
dependence of the differential and total cross section of the process 
$e^+ e^-\to \eta_c \gamma$ the form factor $F(t)$ could be sensitive to the 
whole set of $\psi$-resonances. A suitable parametrisation of the form 
factor, in terms of the contribution of the vector mesons $\rho$, $J/\psi$, 
$\psi(3.77)$, $\psi(4.04)$ and $\psi(4.41)$, can be found in the framework of 
VDM, but with a  $2^4=16$-fold ambiguities in the definition of the signs of the
various constants. The QCD-inspired model for the form factor $F(t)$ on the 
basis of 
the process $c \overline{c}\to \gamma \ell^+ \ell^-$ (with two pole Feynman 
diagrams) results in a parametrisation which is very similar to the 
$J/\psi$-contribution alone. This model 
predicts an asymptotic behavour of the type $|t|^{-1}$ for large $|t|$, in 
contrast with the VDM.
\newpage
\appendix
\section*{}
Here we consider the decays $\eta_c\to \rho^0\rho^0$, $V\to \eta_c\gamma$ and 
$V\to 
\ell^+\ell^-$, whose decay constants are necessary for the calculation of the 
form factor $F(t)$ in VDM.

The matrix element for the process $\eta_c\to \rho^0\rho^0$ can be written as
$${\cal M}=\displaystyle\frac{g(\eta_c\to\rho^0\rho^0)}{m}
\epsilon^{\mu\nu\alpha\beta}U_{1\mu}U_{2\nu}p_{1\alpha}p_{2\beta},$$
where $U_1~(U_2)$ and $p_1~(p_2)$ are the polarisation four-vectors and 
four-momenta of the two produced particles.
Summing over the final polarisations, the partial decay width reads
$$\Gamma(\eta_c\to\rho^0\rho^0)=\displaystyle
\frac{m}{64\pi}|g(\eta_c\to\rho^0\rho^0)|^2
\left(1-\displaystyle\frac{4m_\rho^2}{m^2}\right )^{3/2},$$
the constant $g(\eta_c\to\rho^0\rho^0)$ being dimensionless in the present 
normalisation.

The matrix element of the decay $V\to\ell^+\ell^-$ (see Fig 8, which 
illustrates the notation of the particle four-momenta), can be written as
$${\cal M}={e^2}f_V \overline{u}(k_1)\gamma_{\mu}{u}(k_2)U^{\mu},$$
where $U_{\mu}$  is  the polarisation four-vector of the $V-$meson.  Averaging 
over $U_{\mu}$ and summing over the final 
lepton polarisations, we find, in the limit of zero leptonic mass, 
$$\Gamma(V\to\ell^+\ell^-)=\alpha^2f_V^2\displaystyle\frac{4\pi}{3}m_V.$$

The matrix element of the decay $V\to P\gamma$, where $P$ is a pseudoscalar 
meson  (Fig 9),  can be written as 
$${\cal M}=e\displaystyle\frac{g(V\to P\gamma)}{m_V}
\epsilon^{\mu\nu\alpha\beta}U_{\mu}
e_{\nu}q_{\alpha}k_{\beta}.$$
Summing over the $\gamma$- polarisations  and averaging over the 
$V$-polarisations  yields 
$$\Gamma(V\to P\gamma)=\displaystyle\frac{\alpha g^2(V\to P\gamma)}{24}
m_V\left( 1-\displaystyle\frac{m^2}{m^2_V}\right )^3.
$$
Finally $g_V$ is connected to $g(V\to P\gamma$) and $f_V$ through the 
relation
$$g_V=ef_Vg(V\to P\gamma)m^2_V.$$
We stress that this relation does not fix the sign of the constant $g_V$.


\begin{table*}
\begin{tabular}{|c|c|c|}
Model &$\eta_c\to\gamma e^+e^-$&$\eta_c\to\gamma\mu^+\mu^-$ \\
\hline\hline
VDM&3.99&2.39\\
\hline\hline
QCD, $\Delta=.075$&2.22&0.61\\
QCD, $\Delta=.100$&2.22&0.60\\
QCD, $\Delta=.125$&2.21&0.69\\
\end{tabular}
\caption{Prediction for the coefficients of internal conversion 
in units of $10^{-2}$. We have set $\Delta=2<\vec k ^2_{\perp}/m^2>$.}
\label{tab1}
\end{table*}

\newpage

\newpage

\begin{figure}
\vspace{-4truecm}
\begin{center}

\mbox{\epsfxsize=10.cm\leavevmode\epsffile{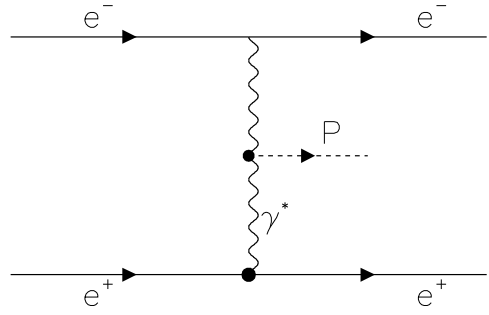}}
\end{center}
\vspace{-5.true cm}
\caption{$2\gamma $-mechanism for the $e^+ e^-\to e^+ e^- P$-process.}
\label{fig1}
\end{figure}

\vspace{-11.true cm}
\begin{figure}
\vspace{9.true cm}
\begin{center}
\mbox{\epsfxsize=12.cm\leavevmode \epsffile{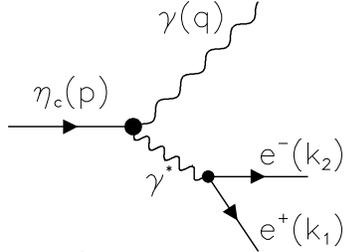}}
\end{center}
\vspace{-12.true cm}
\caption{ Diagram for the process $\eta_c \to \gamma e^+ e^-$, with notation 
of particles four-momenta. }
\label{fig2}
\end{figure}

\ 
\vspace{-5.true cm}
\begin{figure}
\begin{center}
\vspace{-2.true cm}
\mbox{\epsfxsize=12.cm\leavevmode \epsffile{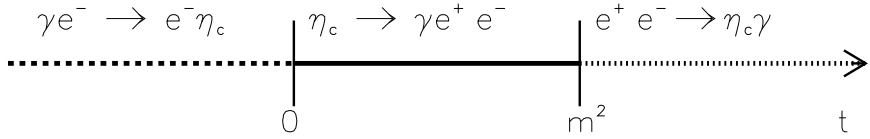}}
\end{center}
\vspace{-8true cm}
\caption{ Scheme for the momentum transfer regions covered by the reactions 
$\gamma e^-\to \eta_c e^- $, $\eta_c\to \gamma e^+ e^-$ and 
$e^+ e^-\to\eta_c \gamma.$}
\label{fig3}
\end{figure}

\newpage
\clearpage
\ 
\begin{figure}
\begin{center}
\vspace{-6.true cm}
\hspace{7.true cm}
\mbox{\epsfxsize=15.cm\leavevmode \epsffile{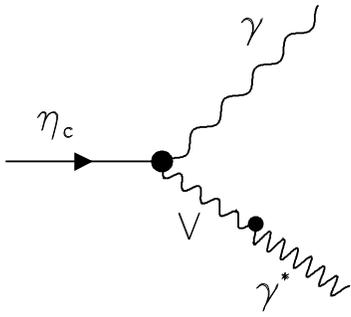}}
\end{center}
\vspace{-14true cm}\caption{ Diagram for the process 
$\eta_c \to \gamma \gamma^*$ in framework of VDM.}
\label{fig4}
\end{figure}

\begin{figure}
\vspace{-3.true cm}
\begin{center}
\mbox{\epsfxsize=12.cm\leavevmode \epsffile{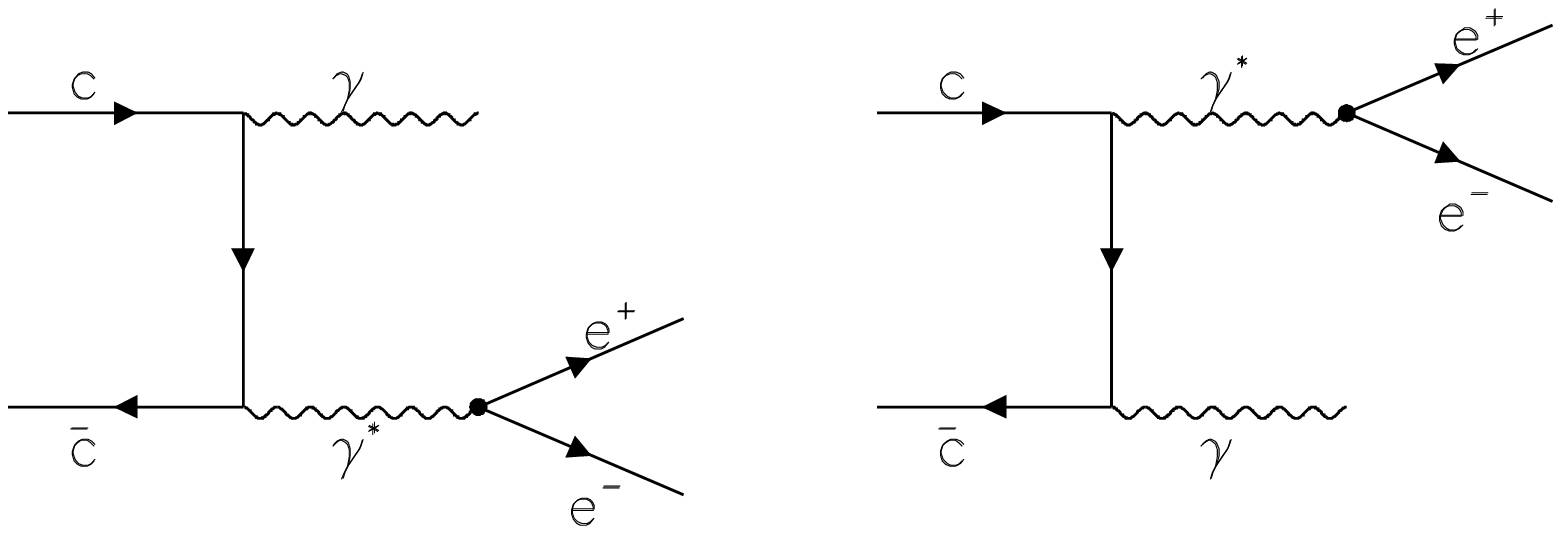}}
\end{center}
\vspace{-10.true cm}
\caption{ Diagrams for the process $c \overline{c}\to \gamma e^+ e^-$.}
\label{fig5}
\end{figure}
\newpage
\clearpage
\vspace{-10.true cm}
\begin{figure}
\vspace{6.true cm}
\begin{center}
\mbox{\epsfxsize=12.cm\leavevmode \epsffile{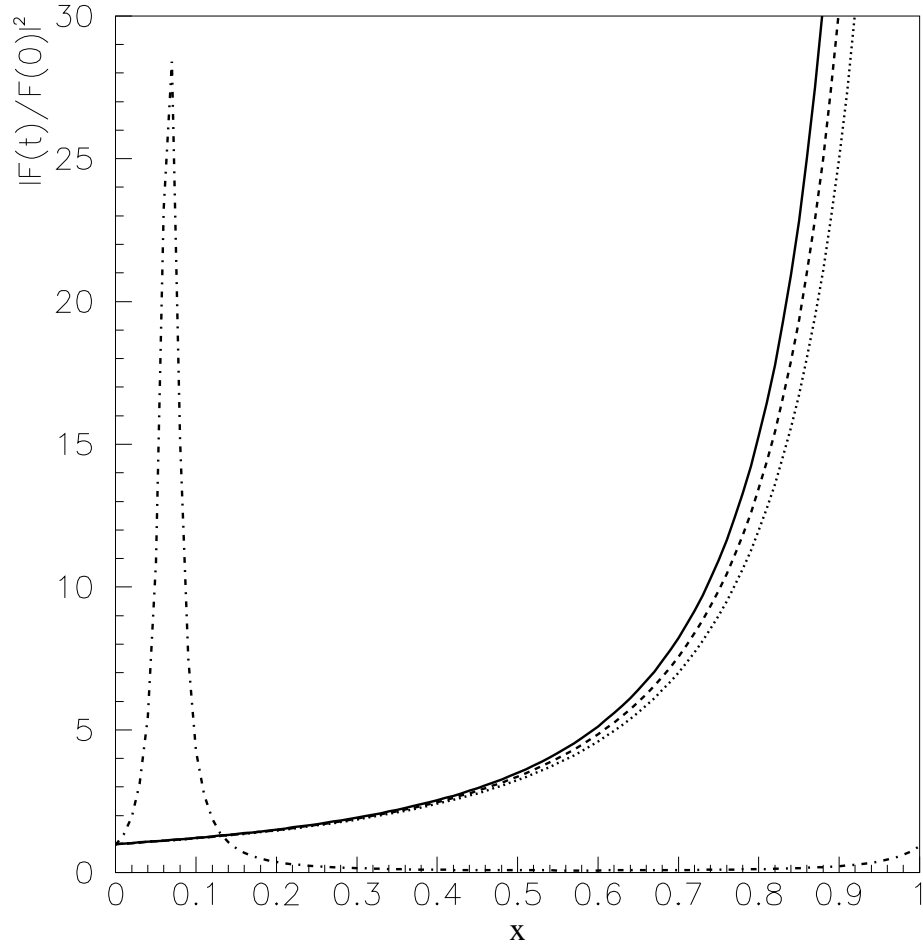}}
\end{center}
\vspace{-1.true cm}
\caption{{\bf x}-dependence ({\bf x} = $t/m^2$) of $|F(t)/F(0)|^2$ in the VDM 
two-pole model 
(dashed-dotted line) and in the QCD-inspired model for different values of the 
transverse quark momentum: $2<\vec k _{\perp}^2 >/m^2=0.075$ (solid line), 
$2<\vec k _{\perp}^2 >/m^2=0.100$ (dashed line), 
and $2<\vec k _{\perp}^2 >/m^2=0.125$ (dotted line).}
\label{fig6}
\end{figure}
\newpage

\vspace{-10.true cm}
\begin{figure}
\vspace{6.true cm}
\begin{center}
\mbox{\epsfxsize=10.cm\leavevmode \epsffile{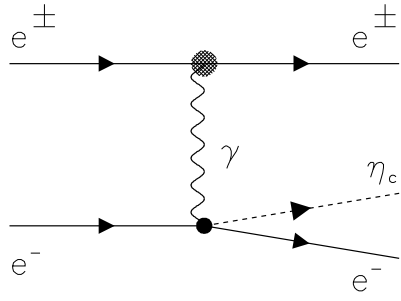}}
\end{center}
\vspace{-6.true cm}
\caption{ Feynman diagram for $e^{\pm} e^- \to e^{\pm} e^- \eta_c $.}
\label{fig7}
\end{figure}

\ 
\begin{figure}
\begin{center}
\vspace{-6.true cm}
\hspace{7.true cm}
\mbox{\epsfxsize=15.cm\leavevmode \epsffile{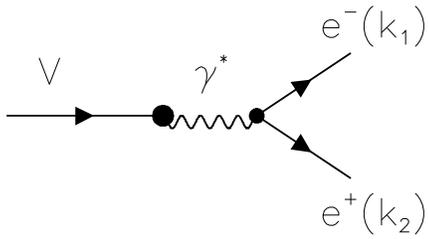}}
\end{center}
\vspace{-14.true cm}
\caption{Diagram of the decay $V\to e^+ e^-$, with notation of particles 
four-momenta, in the framework of one-photon exchange mechanism.}
\label{fig8}
\end{figure}

\ 
\begin{figure}
\begin{center}
\vspace{-6.true cm}
\hspace{7.true cm}
\mbox{\epsfxsize=15.cm\leavevmode \epsffile{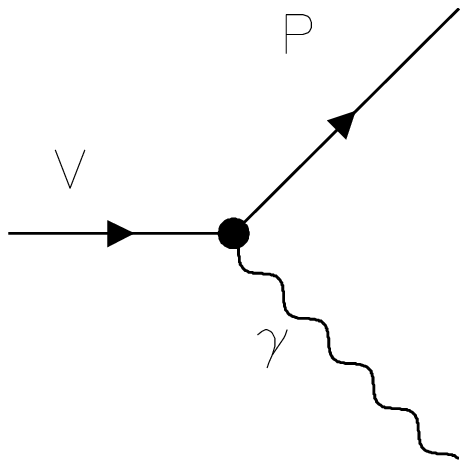}}
\end{center}
\vspace{-14.true cm}
\caption{Diagram of the decay $V\to P \gamma$.}
\label{fig9}
\end{figure}

\begin{thebibliography}{99}
\bibitem{fr} A. Le Yaouanc, Ll. Olivier, O. Pene and J.C. Reynal, {\it Hadron 
Transitions in the Quark Model}, Gordon and Breach Science Publishers, 
London, 1988; Chapt. 2
\bibitem{ap} T. Appelqvist et al., Ann. Rev. Nucl. Part. Sci. {\bf 28}, 
387 (1978)
\bibitem{Fo89} H. Fonvieille et al., Phys. Lett. {\bf B233}, 65 (1989)
\bibitem{Cl95} CLEO collaboration, V. Savinov et al., (1995) hep-ex/9507005
\bibitem{Cl97} CLEO collaboration, J. Gronberg et al., Phys. Rev. {\bf D57}, 33
(1998)
\bibitem{Ja96} R. Jakob, P. Kroll and M. Raulfs, J. Phys. {\bf G22}, 45 (1996)
\bibitem{Kr96}  P. Kroll and M. Raulfs,  Phys. Lett. {\bf B387}, 848 (1996)
\bibitem{Mu97}  I. V. Musatov and A. V. Radyushkin, hep-ph/9702443 (1997)
\bibitem{Bo89}  J. Botts  and G. Sterman, Nucl.  Phys. {\bf B325}, 62 (1989)
\bibitem{L3} L3 collaboration, M. Acciarri et al., Phys. Lett. {\bf B461}, 155 
(1999)
\bibitem{pdg} Particle Data Group, C. Caso et al., EPJ, {\bf C3}, 1 (1998)
\bibitem{svz} V.A. Novikov et al., Phys. Rep. {\bf 41}, 1 (1978) 
\bibitem{rry} L.J. Reinders, H.R. Rubinstein and S. Yazaki, Phys. Rep. 
{\bf 127}, 1 (1985)
\bibitem{ch} K.T. Chao, Nucl. Phys. {\bf B335}, 101 (1990)
\bibitem{rry2} L.J. Reinders, H.R. Rubinstein and S. Yazaki, Phys. Lett. 
{\bf B113}, 414 (1982)
\bibitem{dp} E. Di Salvo and M. Pallavicini, Nucl. Phys. {\bf B427}, 22 (1994)
\bibitem{Al94}  T. M. Aliev, E. Iltan, N. K. Pak, M. P. Rekalo, Z. Phys.
{\bf C54}, 683 (1994)
\bibitem{Du96}  A. Z. Dubnickova, S. Dubnicka and M. P. Rekalo, Z. Phys.
{\bf C70}, 473 (1996)
\bibitem{Kr55}  N. M. Kroll, W. Wada, Phys. Rev. {\bf 98}, 1355 (1955)
\bibitem{Le80} G. P. Lepage and S. J. Brodsky, Phys. Rev. {\bf D22}, 2157 
(1980)
\bibitem{Fe97}  T. Feldmann and P. Kroll, Phys. Lett. {\bf B413}, 410 (1997)
\bibitem{fe}    T. Feldmann, "Fourth International Workshop on Progress in Heavy
Quark Physics", Sept. 20-22, 1997 (hep-ph/9710385); T. Feldmann and P. Kroll, 
"PHOTON '97" {\it Egmond aan Zee, The Netherlands, May 10-15, 1997} 
(hep-ph/9706224)
\bibitem{Wi85}  M. Wirbel, B. Steeh and M. Bauer  Z. Phys. {\bf C29}, 637 (1985)
\end{thebibliography}
\end{document}